\documentstyle[amssymb, prl, aps, epsf, preprint]{revtex}

\begin{document}
\draft
\title{Anomalous acoustoelectric effect in ${\rm La_{0.67}Ca_{0.33}MnO_{3}}$ films}

\author{Y. Ilisavskii,${}^{1}$ A. Goltsev,${}^{1}$ K. Dyakonov,${}^{1,}$\cite{email}
V. Popov,${}^{1}$ E. Yakhkind,${}^{1}$ \\V.P. Dyakonov,${}^{2,3}$
P. Gier{\l}owski,${}^{2}$ A. Klimov,${}^{2}$ S.J.
Lewandowski,${}^{2}$ and H. Szymczak${}^{2}$}

\address{${}^{1}$A.F. Ioffe Physical-Technical Institute, Politekhnicheskaja 26,
St.Petersburg 194021, Russia\\
${}^{2}$Institute of Physics, Al. Lotnik{\'{o}}w 32/46, 02-668
Warszawa, Poland\\
${}^{3}$A.A. Galkin Physico-Technical Institute, R.Luxemburg 72,
Donetsk 340114, Ukraine}

\date{\today}
\maketitle

\begin{abstract}

We have studied acoustoelectric (AE) effect produced by surface
acoustic waves (SAW) in a monolithic layered structure, composed
of piezodielectric LiNbO$_{3}$ substrate and ${\rm
La_{0.67}Ca_{0.33}MnO_{3}}$ film. The experiments unexpectedly
revealed in the longitudinal AE effect an anomalous contribution,
invariant upon reversal of SAW propagation, which coexists with
the ordinary (odd in wave vector) effect. The anomalous effect
dominates near the metal-insulator transition, while the ordinary
effect prevails at high and low temperatures. We show that the
anomalous effect is caused by strong modulation of the film
conductivity produced by the SAW elastic deformations.
\end{abstract}


\narrowtext

The mixed-valence perovskite manganese oxides
R$_{1-x}$A$_{x}$MnO$_{3}$, where R = La, Nd, Pr, and A = Ca, Sr,
Ba, Pb, have been the subject of intense experimental and
theoretical studies over the past years. These compounds attract
much attention not only because they exhibit a rich variety of
strongly interrelated magnetic, structural, and electronic
properties, but also because of possible technical applications.
These studies have shown that the properties of manganites are
determined not only by double-exchange mechanism \cite{Zener} but
also by strong electron-phonon interaction \cite{Millis} of the
Jahn-Teller type. The latter mechanism is responsible for
polaronic states with thermally activated transport in the
paramagnetic state, while the double exchange leads to
ferromagnetic transition at a critical temperature $T_{c}$,
accompanied by a change from semiconductor-like behavior above
$T_{c}$ to metallic-like one below $T_{c}$. There are several
direct experimental observations, which confirm the above outlined
general picture (see e.g. the review by Coey {\it et al.}
\cite{Coey} and references therein). The recently observed large
pressure effect on the transport and magnetic properties of
manganites\cite{P} also is in agreement with the strong
electron-phonon coupling. However, there still remain many open
and unsolved problems.

In this Letter we report on the investigation of the acoustoelectric (AE)
effect in ${\rm La_{0.67}Ca_{0.33}MnO_{3}}$ (LCMO) thin films, which was
undertaken primarily in order to obtain independent experimental evidence
concerning the conduction mechanism and the type of charge carriers in
manganites. We have investigated a monolithic layered structure composed of
piezodielectric LiNbO$_{3}$ (LNO) monocrystalline substrate onto which was
deposited a suitably patterned LCMO thin film. The film was penetrated by a
surface acoustic wave (SAW) launched into the substrate.

The SAW drags free charge carriers due to the momentum transfer from the
acoustic wave to the carriers thus producing in the LCMO film either
acoustoelectric voltage or current, depending on whether the film is open-
or short-circuited. In our structure, charge carriers are affected both by a
deformation wave and piezoelectric field, which accompany the propagating
SAW. This is in contrast to earlier experiments, in which the investigated
sample was subject only to the piezoelectric field \cite{AEE}. Due to the
SAW localization near the substrate-film interface, deformation amplitude in
the film can reach 10$^{-3}$, equivalent to the pressure of 0.1 GPa. This
oscillating pressure causes in turn a modulation of the film conductivity.

We have found that the AE current displays a strong temperature
dependence peaking in the vicinity of the metal-insulator (M-I)
transition. However, the most intriguing result is the observation
of two contributions to the AE current. Besides the ordinary AE
current, which is odd in the SAW wave vector, i.e. it changes its
sign upon reversal of the SAW direction of propagation, we have
observed also an anomalous -- even in the wave vector -- current
component. The sign of the ordinary AE current in the whole
temperature range was found to be in agreement with hole-like
conductivity. The anomalous AE current, as we show further on, is
related to the strong pressure dependence of the manganite film
conductivity. This is supported by the fact that in a different
experimental arrangement only the ordinary AE effect has been
observed \cite{AEE}. The even longitudinal AE effect has been
studied in several situations (e.g. piezoelectrics with traps
\cite{Morozov}, and asymmetric ballistic channels \cite{E-W}), but
these studies clearly are not pertinent in our case. Our manganite
films are spatially homogeneous and contain no traps. Moreover,
geometrical factors have been checked and excluded. Some parallels
can be drawn only to the first theoretical study of the even
acoustoelectric effect, carried out for crystals without the
center of inversion \cite {Gurevich}. Our films have such center,
but are in intimate contact with a substrate lacking this
property. To the role of symmetry we revert further on.

The LCMO films, 100 to 200 nm thick, were laser ablated from
ceramic ${\rm La_{0.67}Ca_{0.33}MnO_{3}}$ target, and grown at
$730 ^{\circ }$C on $+y$ - cut LNO substrates without
post-annealing. After deposition the films were patterned
photolithographically and etched into a Hall bar 10 mm long and 2
mm wide. X-ray diffraction investigation shows that the films are
single phase, epitaxial and (211) oriented with the pseudocubic
lattice parameter $ a_{0}=$ 0.3853 nm. Chemical composition of the
films was found from electron probe microanalysis (EPMA) to be
identical to that of the target, within an experimental error of
2\%. The spatial and electrical homogeneity of the films was
verified by the EPMA scanning and resistive measurements performed
at different parts of the samples.

The AE effect studies have been carried out by launching the
Rayleigh surface acoustic waves along the surface of LNO substrate
(see Fig.~1). The SAW pulse was generated and detected by two
interdigital transducers (IDT), positioned at the edges of the
substrate with the LCMO film deposited centrally between them. The
pulse duration $\tau $ was varied from 1 to 7 $\mu $s and the
pulse repetition rate was kept constant at 50 Hz. The transducer
aperture (3 mm) was larger than the film width and, therefore, the
acoustic wave was spreading over the whole film. During the AE
studies the possible role of spurious effects such as
rectification of stray radio frequency (rf) fields induced by the
IDTs in film contacts, the SAW diffraction and geometrical effects
has been checked and excluded. The AE effect has been studied in
the short-circuit geometry at the SAW frequency of 87 MHz
($\lambda $ = 40 $\mu $m). The AE current between the film
contacts has been determined by measuring the voltage drop across
a load resistor using a video-amplifier and an oscilloscope.

The resistivity $\rho (T)$ of the investigated LCMO films attains a maximum
at 220 K (Fig.~2). According to ac susceptibility measurements employing a
mutual inductance method, the resistivity peak takes place near the
ferromagnetic phase transition (see the inset in Fig.~2). Low values of room
temperature and residual resistivities (20 m$\Omega $cm and $\sim $ 500 $\mu
\Omega $cm at 4.2 K, respectively) and the sharpness of $\rho (T)$ peak
indicate the absence of significant grain boundary contributions to the
resistivity. An applied magnetic field $H$ shifts the $\rho (T)$ peak
towards higher temperatures and markedly increases the film conductivity,
resulting in colossal magnetoresistance effect: MR = $[\rho (H)-\rho
(0)]/\rho (0)\thicksim -80$\% at $25.5$ kOe.

Results of the longitudinal acoustoelectric effect measurements in the LCMO
films are shown in Fig.~3(a), where full circles mark the temperature
dependence of the AE current $I_{AE}$ in zero magnetic field for a sound
wave vector {\bf q} parallel to the $+z$ axis of the substrate (this
direction is distinguished due to the lack of the center of inversion in
LNO). At $T=300$ K the value of $I_{AE}$ is about 2 $\mu $A for the SAW
intensity $\Phi \sim 3$ W/cm. $I_{AE}$ increases with decreasing temperature
and approaches its maximum value of about 25 $\mu $A near the M-I
transition. With further decrease of temperature the AE current is reduced
to $\sim $ 1 $\mu $A at $T\approx 77$ K.

As already remarked, the ordinary longitudinal acoustoelectric effect should
be odd with respect to the SAW wave vector {\bf q} \cite{W}. In order to
check this assertion, we performed separate measurements with reversed
direction of SAW propagation by switching the rf source to the other
transducer, so that {\bf q} was antiparallel to the $+z$ axis. The
acoustoelectric current $I_{AE}$ measured in this manner exhibited an
unexpected and puzzling behavior shown in Fig.~3(b) by full circles. As
seen, $I_{AE}$ displays a complex temperature dependence and undergoes twice
a sign reversal at temperatures near the M-I transition.

Such an unexpected dependence of $I_{AE}$ on the {\bf q} direction
can be explained if two contributions to the AE current are
assumed to exist: $I_{AE}=I_{even}+I_{odd}$. The first term is
anomalous and even in {\bf q}, while the second one is ordinary
and odd. We have separated both these contributions using
experimental data and the relations: $I_{odd}(- {\bf
q})=-I_{odd}({\bf q})$, $I_{even}(-{\bf q})=I_{even}({\bf q})$.
The results are displayed in Fig.~3(a) and (b). One can see that
the anomalous longitudinal AE effect dominates near the M-I
transition and its magnitude exceeds the ordinary longitudinal AE
effect approximately twice. It should be remarked that the
anomalous $I_{AE}$ is always directed along the $+z$ axis. The
ordinary AE effect prevails at high and low temperatures, and its
sign corresponds to the hole-like conductivity in the whole
investigated temperature range, as it is expected in the case of
partial substitution of trivalent La by divalent Ca.

To verify the acoustoelectric nature of the observed effects, we
have investigated in detail the dependence of $I_{AE}$ both on
$\Phi$ and $\tau $. These measurements have shown, as expected,
that at all temperatures $I_{AE}$ depends linearly on $\Phi$,
while the initially linear dependence of $I_{AE}$ on $\tau$
saturates for $\tau>2.5$ $\mu $s, when the spatial length of the
SAW pulse becomes larger than the film length.

To analyze the interaction of the SAW with the thin manganite film
in mechanical contact with the LNO substrate we apply the
coordinate system with the positive $y$ axis normal to the film
(see Fig.~1). The boundary plane is the $xz$ plane. The manganite
film has a thickness $a$, thus the free surface of the film is the
plane $y=a$. The SAW of frequency $\omega $ propagates in the
piezodielectric substrate along the $z$ axis and is accompanied
both by an electric field ${\bf E}(y,z,t)$ and strain
$S_{ij}(y,z,t)$. In our geometry, only the $y$ and $z$ components
of ${\bf E}$ and the lattice displacement are nonzero. ${\bf E}$
generates in turn a local current density $J_{i}(y,z,t)=\sigma
_{ij}(y,z,t)E_{j}(y,z,t)$ in the film, where $\sigma _{ij}$ is the
tensor of the film conductivity (in the long-wavelength limit the
diffusive contribution to the current may be omitted). At the
acoustic frequencies used ($\omega \thicksim 10^{8}$ s$^{-1} $),
the dependence of $\sigma $ on $\omega $ may be neglected, as the
strong frequency dependence of the polaronic conductivity is
expected to occur at $\omega \thicksim (4/\hbar ){\mathbb
E}_{a}\thicksim 10^{13}$ s$^{-1}$ (see e.g. \cite {Reik}), where
${\mathbb E}_{a}\thicksim 120$ meV is the polaronic activation
energy in LCMO.

${\bf E}$ and $S_{ij}$ induce a local modulation of $\sigma _{ij}$
: $\sigma _{zz}(y,z,t)=\sigma _{0}+\sigma _{1}(y,z,t)+\sigma
_{2}(y,z,t)$, where $\sigma _{0}$ is the unperturbed conductivity,
$\sigma _{1}(y,z,t)=n_{s}(y,z,t)\partial \sigma _{0}/\partial n$ ,
representing the effect of electric field ${\bf E}$, is due to the
modulation of charge carrier concentration $n=n_{0}+n_{s}$,
$n_{s}\ll n_{0}$ by ${\bf E}$ {\bf (} the influence of ${\bf E}$\
on the drift mobility may be neglected{\bf )}, and the last term
\[
\sigma _{2}(y,z,t)=\sigma _{0}[\Pi _{3333}S_{zz}+\Pi _{3322}S_{yy}+\Pi
_{3323}S_{yz}]
\]
describes the modulation of $\sigma _{zz}$ by the strain. The
tensor $\Pi _{ijkl}\equiv \partial \ln \sigma _{ij}/\partial
S_{kl}$ calculated at $ S_{kl}=0$ and at fixed temperature
describes the effect of strains on $\sigma _{ij}$. The changes of
drift mobility and electron concentration both contribute to $\Pi
_{ijkl}$. The pseudocubic symmetry of LCMO allows to put $\Pi
_{3323}=0$, and for the other two terms we will simplify the
notation to $\Pi _{3333}\equiv \Pi _{33}$, and $\Pi _{3322}\equiv
\Pi _{32}$.

The longitudinal AE current per unit length in $x$ direction generated by
the SAW in the film is by definition
\begin{equation}
j_{ae} = \frac{1}{\Theta }\int_{0}^{\Theta }dt\int_{0}^{a}dyJ_{z}(y,z,t)
=\frac{1}{\Theta }\int_{0}^{\Theta }dt\int_{0}^{a}dy{(\sigma
_{1}} E_{z}+\sigma _{2}E_{z})\equiv j_{ae}^{(1)}+j_{ae}^{(2)},
\label{aee}
\end{equation}
where $\Theta =2\pi /\omega $. In order to calculate $j_{ae}$ it
is necessary to find the relations between $E_{z}$, $S_{ij}$ and
$n_{s}$, imposed by Maxwell's equations and
mechanical-piezoelectric equations of state, and by the boundary
conditions at the surfaces $y=0$ and $y=a$. The thickness $a$ of
our film is small compared to the acoustic wavelength $\lambda $,
but still much larger than the Debye length $\lambda _{D}$, i.e.,
$\lambda _{D}\ll a\ll \lambda $. In the broad temperature region
around the $\rho (T)$ peak, the conductivity of the manganite film
is small, close to that of semiconductors. Therefore, we can apply
the Ingebrigtsen's approach \cite{Ingebrigtsen70}, and assume that
the electric field of the SAW produces at the film surfaces $y=0$
and $y=a$ surface charges due to the screening effect; these
charges produce in turn surface currents. The dc component of the
total surface current per unit length in $x$ direction is the
current $j_{ae}^{(1)}$:
\begin{equation}
j_{ae}^{(1)}=q\Gamma \Phi {\sigma _{0}}(e\omega n{_{0}})^{-1},  \label{ae1w}
\end{equation}
where $e$ is the charge of charge carriers, $\Phi $ is the SAW intensity,
and $\Gamma $ is the attenuation ($\Phi =\Phi _{0}\exp (-\Gamma z)$):
\begin{equation}
\Gamma =\frac{2\pi }{\lambda }K^{2}\frac{\sigma _{\Box }/\sigma
_{m}}{1+(\sigma _{\Box }/\sigma _{m})^{2}},  \label{att}
\end{equation}
where $K^{2}$ is the electromechanical coupling coefficient, $\sigma _{\Box
}=a\sigma _{0}$ is the sheet conductivity, $\sigma _{m}$ is a material
constant \cite{Ingebrigtsen70}. $j_{ae}^{(1)}$ is odd in $q$ and represents
the ordinary longitudinal AE current \cite{W,note}. Eq.~(\ref{att})
describes very well the observed SAW attenuation (Fig.~2). A noticeable
deviation of the experimental dependence $\Gamma (T)$ from Eq.~(\ref{att})
occurs only at temperatures approximately 30 K below the M-I transition.

In order to calculate the current $j_{ae}^{(2)}$ due to the film
deformation, it is necessary to evaluate the quantities $S_{ij}$
and $E_{z}$ in the film, taking into account their continuity at
$y=0$. Since the film is thin ($aq\ll 1$), we can assume that
$S_{zz}$ changes weakly within $ 0<y<a $, i.e.
$S_{zz}(y,z,t)\approx S_{zz}(0,z,t)$. The strain $S_{yy}$ may be
estimated as $S_{yy}(y,z,t)\approx -\nu S_{zz}(0,z,t)$, which is
exact at the free surface $y=a$ and $\nu =c_{12}/c_{11}\approx
0.4$\cite{Poisson} where $c_{11}$ and $c_{12}$ are the components
of the elastic tensor of the LCMO film. In LNO the strain $S_{zz}$
is related to the electrical displacement $D_{z}$ and electric
field $E_{z}$ by $D_{z}=\varepsilon _{33}E_{z}+p_{33}S_{zz}$,
where $\varepsilon _{33}$ and $p_{33}$ are components of the
dielectric and piezodielectric tensor, respectively (we assumed
$p_{32}=0$, as this parameter weighs about 1\% in our results). In
this manner, Eq.~(\ref{aee}) gives
\begin{equation}
j_{ae}^{(2)} =\frac{a\sigma _{0}}{p_{33}}(\Pi _{33}-\nu \Pi
_{32})\frac{1}{
\Theta }\int_{0}^{\Theta }dt[D_{z}(+0,z,t)  
-\varepsilon _{33}E_{z}(0,z,t)]\overline{E}_{z}(z,t),  \label{ae2}
\end{equation}
where $\overline{E}_{z}(z,t)$ is the electric field averaged over
the film thickness. For $aq\ll 1$ we have
$\overline{E}_{z}(z,t)\approx E_{z}(0,z,t)$. We estimated the
contribution of $D_{z}$ to $j_{ae}^{(2)}$ in Eq.~(\ref {ae2}), and
found that $D_{z}$ can be neglected if $1\ll \sigma
_{0}/\varepsilon \omega \ll 1/(\lambda _{D}q)$, where $\varepsilon
$ is the static dielectric constant of the film. Using the
relation $ E_{z}(0)=iq\varphi (0)$ and expressing the surface
potential $\varphi (0)$ in terms of $\Phi $ and $\Gamma $, we
obtain from Eq.~(\ref{ae2})
\begin{equation}
j_{ae}^{(2)}=-\Gamma \Phi \varepsilon _{33}(\Pi _{33}-\nu \Pi _{32})/p_{33}.
\label{ae2f}
\end{equation}
This result clearly demonstrates that $j_{ae}^{(2)}$ is even in $q$, and its
direction is determined by the signs of $p_{33}$ and $\Pi _{3i}$. It is
interesting to note that the current $j_{ae}^{(2)}$ is bulk, unlike the
surface current $j_{ae}^{(1)}$. The coefficients $\Pi _{3i}$ are related to
the pressure dependence of the conductivity: $2\Pi _{32}+\Pi _{33}=-3\kappa
^{-1}\partial \ln \sigma _{0}/\partial P$, where $\kappa $ is the film
compressibility. According to the pressure experiments, in LCMO \cite{P} the
quantity $\partial \ln \sigma _{0}/\partial P$ is positive, which means
negative $\Pi _{3i}$, and has a pronounced temperature dependence: it is
small at high and low temperatures but peaks to about 3.5 [GPa]$^{-1}$ at a
temperature slightly lower than that of the resistivity peak. For numerical
estimation we will approximate $\Pi _{33}\approx \Pi _{32}$. Using $\kappa
^{-1}=85$ GPa \cite{Zhu} we find max$\Pi _{ij}\sim -300$.

It should be observed from Eq.~(\ref{ae2f}) that the appearance of
the even AE effect requires a pressure dependent film conductivity
($\Pi _{3i}\neq 0)$, piezoelectric properties of the substrate
($p_{33}\neq 0$), and the existence of a distinguished direction
in the substrate.

The above outlined theory is in a good agreement with the
experimental data. As the coordinate $z$ axis is chosen to be
parallel to the $+z$ crystallographic axis, the constant $p_{33}$
is positive \cite{ep} and, according to Eq.~(\ref{ae2f}),
$j_{ae}^{(2)}$ flows along $+z$ axis. Substituting the well known
parameters of LNO and our own experimental result max$\Gamma \sim
2$ cm$^{-1}$ into Eq.~(\ref{ae2f}), we find max$j_{ae}^{(2)} \sim
30$ $\mu$A/cm for $\Phi \sim 3$ W/cm. The experimental data in
Fig.~3 correspond to max$j_{ae}^{(2)} \sim 100$ $\mu$A/cm. Taking
into account the approximate character of our estimations of
$\Gamma$, $\Phi$ and $\Pi _{3i}$ we consider the agreement between
the theory and experiment as satisfactory. Eq.~(\ref{ae2f}) also
describes very well the temperature behavior of $j_{ae}^{(2)}$, as
shown in Fig.~3(b). The theoretical curve shown there, normalized
to the experimental maximum, was calculated from Eq.~(\ref{ae2f})
using experimental dependencies $\Gamma(T)$ (Fig.~2) and $\partial
\ln \sigma _{0}/\partial P$ from pressure measurements \cite{P}.
The observed maximum of $j_{ae}^{(2)}$ is shifted towards lower
temperatures by approximately 15 K with respect to the theoretical
prediction. This shift may be caused by a possible difference in
temperature of the peak of $\partial \ln \sigma _{0}/\partial P$
between our LCMO films and bulk samples \cite{P}. It should be
noted that also the order of magnitude of the odd AE current
$j_{ae}^{(1)}$ estimated from Eq.~(\ref{ae1w}) agrees with the
experiment.

In conclusion, we have observed two contributions to the
longitudinal acoustoelectric (AE) effect produced by surface
acoustic waves in La$_{0.67}$Ca$_{0.33}$MnO$_{3}$ films. The
anomalous, or even in the acoustic wave vector, AE effect is
dominant near metal-insulator transition, where it exceeds a few
times the ordinary (odd) AE effect. The ordinary AE effect
prevails at high and low temperatures, and its sign corresponds to
hole-like conductivity. The anomalous AE effect is shown to be due
to a strong modulation of the film conductivity produced by
elastic deformations carried by the acoustic wave. Such effect can
be expected also in other conducting materials, which exhibit
similar pressure dependence of conductivity.

This work was supported in part by Polish Government (KBN) Grant 2 P03B 139
18 and PBZ-KBN-013/T08/19, and the Russian Foundation for Basic Research
Grants 99-02-18333 and 01-02-17479.


\newpage

\begin{figure}[tbp]
\centering
   \leavevmode
   \epsfclipon
   \epsfbox{
         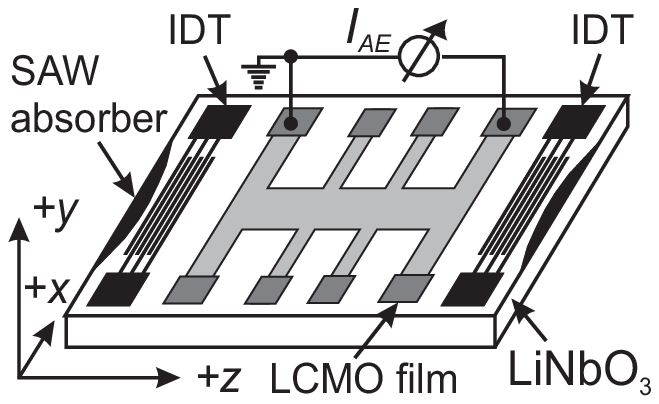}
   \vspace{5 mm}
\caption{ Schematic drawing of the investigated structure showing the LiNbO$%
_{3}$ substrate, La$_{0.67}$Ca$_{0.33}$MnO$_{3}$ (LCMO) film deposited on
the substrate, and interdigital transducers (IDT). The coordinate system
refers to the LiNbO$_{3}$ crystallographic axes. }
\label{fig1}
\end{figure}

\newpage

\begin{figure}[tbp]
\centering
   \leavevmode
   \epsfclipon
   \epsfbox{
         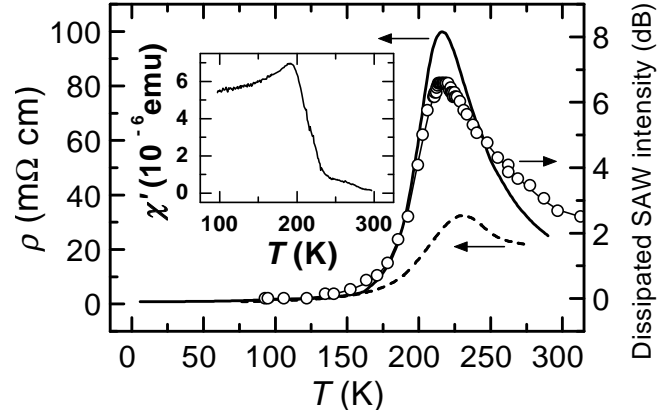}
   \vspace{5 mm}
\caption{ Sample resistivity versus temperature at $H=0$ (solid line) and $%
H=25.5$ kOe (dashed line), and SAW dissipation in the LCMO-LNO layered
structure at $H=0$ (open circles). Inset: ac susceptibility measured using
in-plane ac magnetic field of 5 Oe at 625 kHz versus temperature. }
\label{fig2}
\end{figure}

\newpage

\begin{figure}[tbp]
\centering
   \leavevmode
   \epsfclipon
   \epsfbox{
         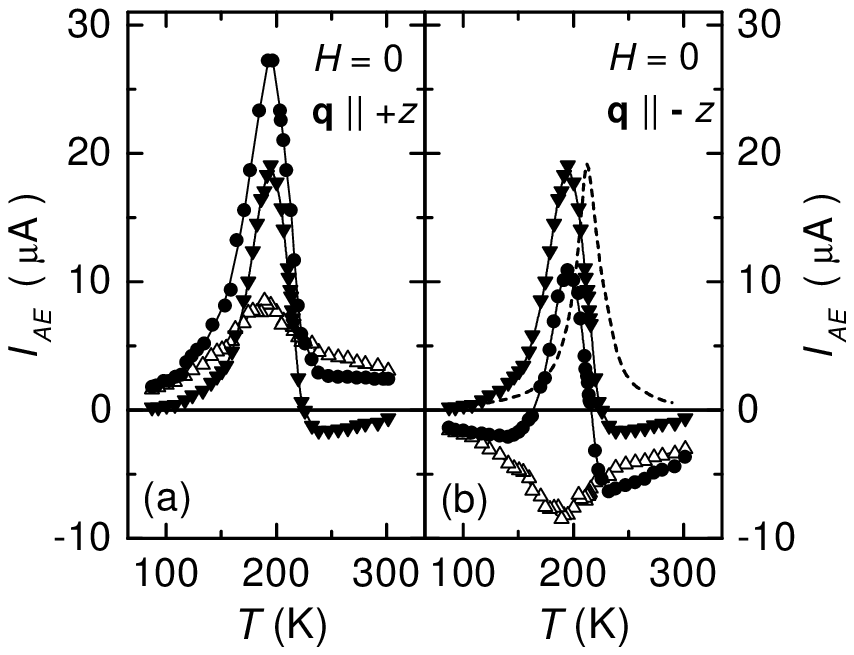}
   \vspace{5 mm}
\caption{ Acoustoelectric current $I_{AE}$ versus temperature in La$_{0.67}$%
Ca$_{0.33}$MnO$_{3}$ film at $H=0$ for two orientations of the SAW wave
vector {\bf q} with respect to $+z$ axis of LiNbO$_{3}$: (a) parallel and
(b) antiparallel. $\bullet$ - total AE current; $\blacktriangledown$ - the
anomalous (even) contribution to $I_{AE}$; $\vartriangle$ - the ordinary
(odd) contribution to $I_{AE}$; ${\bf -}$\,${\bf -}$\,${\bf -}$ \, - the
even AE current calculated from Eq.~(\ref{ae2f}). The solid lines are a
guide for the eye. }
\label{fig3}
\end{figure}


\begin{references}
\bibitem[{*}]{email}  Electronic address: K.Dyakonov@pop.ioffe.rssi.ru

\bibitem{Zener}  C. Zener, Phys. Rev. {\bf 82}, 403 (1951).

\bibitem{Millis}  A.J. Millis, P.B. Littlewood, and B.I. Shraiman,
Phys.Rev.Lett. {\bf 74}, 5144 (1995).

\bibitem{Coey}  J.M.D. Coey, M. Viret, and S. von Moln\'{a}r, Adv.Phys. {\bf %
48}, 167 (1999).

\bibitem{P}  J.J. Neumeier {\it et al.}, Phys.Rev. B {\bf 52}, R7006 (1995).

\bibitem{AEE}  Li Wang {\it et al.}, Phys.Rev. B {\bf 60}, R6976 (1999).

\bibitem{Morozov}  A.I. Morozov, JETP Lett. {\bf 2}, 228 (1965).

\bibitem{E-W}  O. Entin-Wohlman {\it et al.}, Phys.Rev. B {\bf 62}, 7283
(2000).

\bibitem{Gurevich}  V.L. Gurevich and A.L. Efros, Sov.Phys.JETP {\bf 17},
1432 (1963); V.L. Gurevich, Sov.Phys.Semicond. {\bf 2}, 1299 (1969).

\bibitem{W}  G. Weinreich, Phys.Rev. {\bf 107}, 317 (1957).

\bibitem{Reik}  H.G. Reik, Solid State Commun. {\bf 1}, 67 (1963).

\bibitem{Ingebrigtsen70}  K.A. Ingebrigtsen, J.Appl.Phys. {\bf 41}, 454
(1970).

\bibitem{note}  The deformation potential interaction of charge carriers
with the SAW gives an additional contribution to the odd AE current. In
manganites this term can be neglected, because even at very low temperatures
$\lambda $ is much larger than the mean free path of charge carriers.

\bibitem{Poisson}  C. Zhu and R. Zheng, J.Appl.Phys. {\bf 87}, 3579 (2000).

\bibitem{Zhu}  C. Zhu {\it et al.} Appl.Phys.Lett. {\bf 74}, 3504 (1999).

\bibitem{ep}  R.T. Smith and F.S. Welsh, J.Appl.Phys. {\bf 42}, 2219 (1971);
R.A. Graham, J.Appl.Phys. {\bf 48}, 2153 (1977).
\end{references}
\end{document}